\documentclass[twocolumn,runningheads]{svjour2}
\smartqed  % flush right qed marks, e.g. at end of proof
\usepackage{graphicx}
\journalname{Astrophysics and Space Science}
\def\be{\begin{equation}}
\def\ee{\end{equation}}
\def\lsim{\lower 2pt \hbox{$\, \buildrel {\scriptstyle <}\over
         {\scriptstyle \sim}\,$}}
\def\gsim{\lower 2pt \hbox{$\, \buildrel {\scriptstyle >}\over
         {\scriptstyle \sim}\,$}}

\begin{document}

\title{The Geminga Fraction%\thanks{Grants or other notes
}
\subtitle{}


\author{Alice K. Harding         \and
        Isabelle A. Grenier \and
        Peter L. Gonthier}


\institute{Alice K. Harding \at
              Code 663, NASA Goddard Space Flight Center \\
              Greenbelt, MD 20771, USA \\
              Tel.: 301-286-7824\\
              Fax: 301-286-1682\\
              \email{harding@twinkie.gsfc.nasa.gov}    
           \and
           Isabelle A. Grenier \at
              AIM, Service d'Astrophysique, CEA Saclay \\
              91191 Gif Sur Yvette, France
           \and
            Peter L. Gonthier \at
              Dept. of Physics, Hope College \\
              27 Graves Place, Holland, MI 49423, USA 
}

\date{Received: date / Accepted: date}
           
\maketitle

\begin{abstract}
Radio-quiet $\gamma$-ray pulsars like Geminga may account for a number of the unidentified EGRET
sources in the Galaxy.  The number of Geminga-like pulsars is very sensitive to the geometry of 
both the $\gamma$-ray and radio beams.
Recent studies of the shape and polarization of pulse profiles of young radio pulsars have provided evidence 
that their radio emission originates in wide cone beams at altitudes that are a significant fraction (1 -10\%) 
of their light cylinder radius.  Such wide radio emission beams will be visible at a much larger 
range of observer angles than the narrow core components thought to originate at lower altitude.  
Using 3D geometrical modeling that includes relativistic effects from pulsar rotation, we study the 
visibility of such 
radio cone beams as well as that of the $\gamma$-ray beams predicted by slot gap and outer gap models.  
From the results of this study, one can obtain revised predictions for the fraction of Geminga-like, radio 
quiet pulsars present in the $\gamma$-ray pulsar population.
\end{abstract}

\section{Introduction}
\label{sec:intro}

Rotation-powered pulsars and their winds are presently the only known Galactic gamma-ray sources emitting at 
energies above 100 MeV.
The EGRET detector on the Compton Gamma-Ray Observatory (CGRO) detected six $\gamma$-ray pulsars with high confidence\cite{Thom04}, five of which were known radio pulsars.  The sixth, Geminga, is 
radio quiet (or at least an extremely weak radio source\cite{KL99}).  In addition to detecting many sources with 
known counterparts at other wavelengths, EGRET detected some 110 sources\cite{CG06} with no known counterparts, about 
a third of which seem to be of Galactic origin.  It is quite plausible that a sizeable fraction of these 
unidentified EGRET sources are radio-quiet or radio-weak $\gamma$-ray pulsars, i.e. Gemingas.  There are
two possible reasons that may cause a $\gamma$-ray pulsar to be radio-quiet.  The $\gamma$-ray emission may 
have a wider beam than the radio beam or be oriented in a different direction, or the radio emission may be 
either absent or
too weak to be detectable by current radio telescopes.  Geminga-like pulsars are therefore very likely
candidates for some of the unidentified EGRET sources in the Galactic plane.  

It has proven difficult to discover Geminga-like pulsars since EGRET typically did not
collect enough photons in a source to allow pulsation searches without an ephemeris known from other
wavelengths.  Geminga itself, the second brightest EGRET source, 
was only identified as a $\gamma$-ray pulsar after its period was discovered
by ROSAT in the X-ray band\cite{hh92}, although a direct detection in EGRET data was later found to be 
feasible\cite{Bert92}.  It is therefore useful to estimate the number
of Gemingas expected through modeling of the $\gamma$-ray pulsar population assuming different radio and
$\gamma$-ray emission models.  A number of studies of this type have been done (\cite{Gon02,Gon04} 
for the polar cap model and \cite{CZ98} and \cite{JZ06} for the outer gap model).  The results of
these studies have shown that the Geminga fraction, or the number of radio-quiet $\gamma$-ray pulsars relative to the total detectable number of $\gamma$-ray pulsars, strongly depends on the $\gamma$-ray emission
model.  

There are two main types of $\gamma$-ray emission models for which population studies have been carried
out.  Polar cap models assume that particle acceleration takes place near the neutron star magnetic poles and
that $\gamma$-ray emission results from cascades initiated by magnetic pair production in strong magnetic
fields\cite{dh82}\cite{dh96}.  Outer gap models assume that the particle acceleration takes place 
in vacuum gaps that form in the outer magnetosphere\cite{CHR86} and that $\gamma$ rays result from 
cascades initiated by pair production of $\gamma$ rays and soft X-ray photons from the neutron star 
polar cap\cite{cheng94}\cite{Rom96}.  The population studies that assume these two types of models show that 
the Geminga fraction in outer gap models is much higher than in polar cap models.  However, most of these
studies\cite{Gon02}\cite{CZ98} have assumed very simplified geometry for the radio and 
$\gamma$-ray emission beams, taking constant solid angles for the radio and $\gamma$-ray beams, or 
assuming random and independent relative orientations for these beams.  Gonthier et al.\cite{Gon04}\cite{Gon06b} have more recently explored more complex and realistic emission models for the radio and
polar-cap/slot-gap $\gamma$-ray beams.  Jiang et al.\cite{JZ06} have studied the solid angle
dependence of outer gap $\gamma$-ray emission beams in their population model, but did not consider
the relative orientations of the radio and $\gamma$-ray beams.  Furthermore, there has never been a
single study that compares polar-cap/slot-gap models with outer gap
models on the same footing, using an identical set of assumptions for the neutron star population, 
its evolution and 
radio emission geometry, taking into account the relative orientations of the radio and $\gamma$-ray beams. 

The description of the radio emission in these studies has necessarily been empirical, since a physical
model for the coherent radio emission does not exist.  The empirical models that have been developed 
from study of radio pulse profile morphology\cite{Rankin93}\cite{KG03}\cite{ACC02} have 
indicated that the radio emission consists of core emission along the magnetic pole and one or more 
wider cones of emission.  The study of Arzoumanian et al.\cite{ACC02}, as well as some earlier 
studies\cite{Rankin90}, concluded that shorter period pulsars
had more dominant core emission.  Yet, several recent studies of the pulse polarization of young pulsars\cite{CK03}\cite{JW06} and of pulsars showing three peaks in their profiles\cite{Gon06a}, 
suggest that pulsars with short periods, likely $\gamma$-ray pulsars, have more dominant cone beams, implying
a much wider beam of radio emission that in  previous models.

This paper will present the results of a study of both the slot gap and outer gap
models, using the same population code and incorporating the revised radio emission model.  We 
begin by reviewing the most widely studied models of pulsar high-energy emission, the polar cap and its extension to
the slot gap, and the outer gap with its more recent extensions and modifications.  We then discuss
the traditional and revised models for radio emission.  The neutron star population synthesis is
briefly discussed, since it is presented in more detail by Gonthier et al.\cite{Gon06b} 
in these proceedings.
The preliminary results of our study comparing high-altitude slot gap and outer gap models are then 
presented.  Together with the study of the polar cap/low-altitude slot gap presented in 
Gonthier et al.\cite{Gon06b}
using the same set of evolved neutron stars, we can draw some important conclusions about the Geminga
fraction in different models as well as the effect of radio and $\gamma$-ray emission geometry on the
numbers of radio-loud and radio-quiet $\gamma$-ray pulsars.
 
\section{Gamma-Ray Emission Models}
\label{sec:grmods}

Although the high-energy emission from rotation-powered pulsars has been studied for three decades, the
particle acceleration and the location and mechanism of the emission is still not understood.  All models
involve electrostatic acceleration by an electric field parallel to the magnetic field, but in different
regions of the magnetosphere.  All models also involve the production of electron-positron pairs and the
cascades that determine the geometry of the observed $\gamma$-ray emission, by means of curvature and 
synchrotron radiation.

\subsection{Polar cap and slot gap}
\label{sec:SG}

In polar cap models, particle acceleration develops along open magnetic field lines above the neutron star 
surface.  The accelerators divide into two types that depend on how charge is supplied and distributed.  
The two main subclasses are vacuum gap models \cite{ruderman75}, where charges are trapped in
the neutron star surface layers by binding forces and a region of vacuum forms above the surface, and 
space-charge limited flow (SCLF) models \cite{as79}, where charges are freely emitted
from the surface layers.  In SCLF accelerators, a voltage develops due to the small
charge deficit between the real charge density $\rho$ and the Goldreich-Julian charge density 
$\rho_{GJ}  \simeq - 2\epsilon_0 \vec{\Omega}
\cdot \vec{B}$ ($\vec{\nabla} \cdot \vec{E_{\parallel}} = (\rho - \rho
_{GJ})/\epsilon_0$), due to the curvature of the field 
\cite{as79} and to general relativistic inertial frame dragging \cite{mt92}, where $\Omega$ and $B$ are
the rotation rate and surface magnetic field.  
Accelerated particles radiate $\gamma$-rays that create electron-positron pairs in the intense magnetic field
and the $E_{\parallel}$ is screened above a pair formation front (PFF) by polarization of the pairs.  
The potential drop is thus self-adjusted to give particle Lorentz factors around $10^7$, 
assuming a dipole magnetic field.
Above the PFF, force-free conditions could develop if the pair multiplicity is sufficient.  This is likely
for relatively young pulsars, which can produce pairs through curvature radiation\cite{HM01}, 
but older pulsars that can
produce pairs only through inverse-Compton scattering are expected to be pair starved 
and their open magnetospheres
would not achieve a force-free state\cite{HM02,MH04b}.
Given the small radiation loss length scales for particles of these energies, the high energy radiation will
occur within several stellar radii of the surface.  The radiation from electromagnetic cascades produces a 
hollow cone of emission 
around the magnetic pole, with opening angle determined by the polar cap half-angle, 
$\theta_{_{PC}}(r) \sim (2\pi r/Pc)^{1/2}$, at the radius of emission $r$.

More recent versions of the polar cap model\cite{MH03}\cite{MH04a} have explored acceleration in the
`slot gap' (Figure 1), a narrow region bordering the last open field line in which the electric field is 
unscreened\cite{Arons83}. 
Near the open field line boundary, where the electric field vanishes, a larger distance is required for the electrons to accelerate to the Lorentz factor needed to radiate photons energetic enough to produce pairs.  The PFF thus occurs at higher altitudes as the boundary is approached and curves upward, approaching infinity and becoming asymptotically parallel to the last open field line.  If the electric field is effectively screened above the PFF, then a narrow slot surrounded by two conducting walls is formed.   Pair cascades therefore do not take place near the neutron star surface in the slot gap, as do the pair cascades along field lines closer to the magnetic pole (core), but occur on the inner edge of the slot gap at altitudes of 
several stellar radii\cite{MH03}.  The high-energy emission beam of the slot gap cascade is a
hollow cone with much larger opening angle than that of the polar cap cascade emission.  Even so, small values
of both $\alpha$ and $\zeta$, the angle of the magnetic axis and observer direction to the rotation axis 
respectively, (about $10^{\circ}$) are required to reproduce double-peaked profiles of observed $\gamma$-ray pulsars.

\begin{figure}
\hskip -3.0cm
\includegraphics{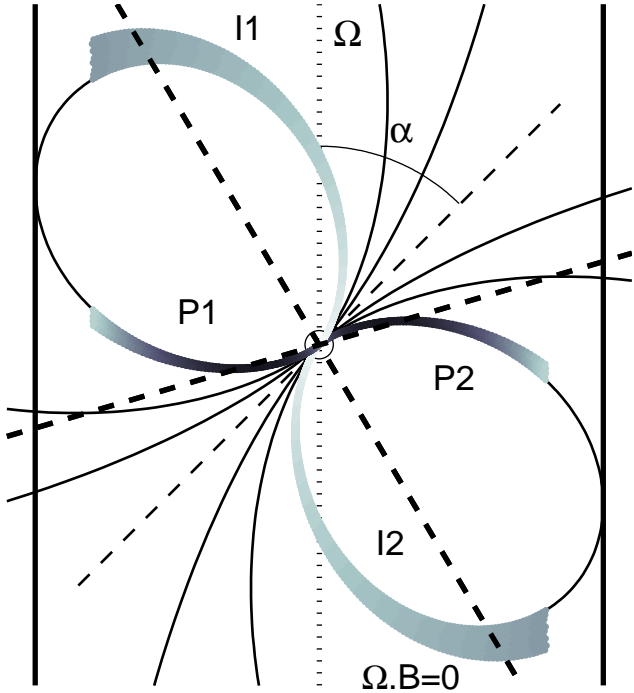}
\caption{2D cross section of magnetosphere showing the location of the extended slot gap accelerator,
for inclination angle $\alpha = 45^\circ$.}
\label{fig:1}       % Give a unique label
\end{figure}

Emission is also expected to occur from the primary particles that continue to accelerate in the slot gap 
at high altitudes because the potential in the slot gap is unscreened.  
Muslimov \& Harding\cite{MH04a} have modeled the acceleration and emission
pattern in the extended slot gap, finding that the electrons reach and maintain Lorentz factors $\sim 10^7$.
Such emission will include curvature, inverse Compton and possibly synchrotron radiation. 
The emission pattern of the high-altitude slot gap radiation forms caustics along the trailing field lines and
displays the geometry of the two-pole caustic model studied by Dyks \& Rudak\cite{dr03}.
The caustic pattern, as seen in Figure \ref{fig:3}, results from a near-cancellation of phase shifts due to aberration, time-of-flight
and curvature of the dipole field along the last open lines on the trailing side\cite{Morini83}.

\subsection{Outer gap}
\label{sec:OG}

\begin{figure}
%\vskip 9cm
\hskip -3.5cm
\includegraphics{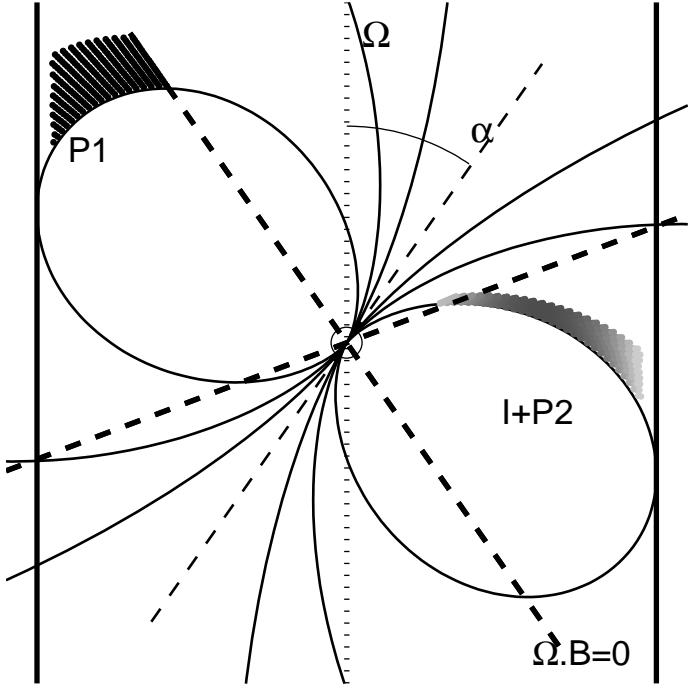}
\caption{2D cross section of magnetosphere showing the location of the outer gap accelerator,
for inclination angle $\alpha = 35^\circ$.}
\label{fig:2}       % Give a unique label
\end{figure}

Outer gap models \cite{CHR86,Rom96} focus on
regions in the outer magnetosphere that cannot fill with charge,
since they lie along open field lines crossing the null surface,
$\vec \Omega \cdot \vec B = 0$, where $\rho _{GJ}$ reverses sign (see Figure \ref{fig:2}).
Charges pulled from the polar cap, along field lines with little or
no pair cascading from the polar cap accelerator, therefore cannot populate the
region between the null surface and the light cylinder, and a
vacuum gap forms. If outer gaps form, they can accelerate particles to high
energy and the radiated $\gamma$ rays can produce pairs by
interacting with thermal X-rays from the neutron star surface. 
Although the
density of such X-ray photons is very small in the outer gaps, it
is enough to initiate pair cascades since the newborn pairs
accelerate in the gap, radiate, and produce more pairs.  The gap
size is limited by the pair cascades, which screen the gap
electric field both along and across field lines, thus determining
the emission geometry. Young pulsars, having hotter polar caps and
higher vacuum electric fields, tend to have narrow gaps stretching
from near the null surface to near the light cylinder
\cite{cheng94} while the gaps of older pulsars, having lower
electric fields, are much thicker and grow with age
\cite{zhang97}. When the gap fills the whole outer magnetosphere
(at ages $\tau \gsim 10^7$ yr) it ceases to operate, so that not all
radio pulsars can emit $\gamma$ rays. Death lines in $P$-$\dot P$
space predict which pulsars can sustain outer gaps, depending on
whether the X-ray photon field comes from cooling of the whole
stellar surface or from polar caps heated by the energy deposited
by the return flux of charges \cite{zhang04}.

The recent outer gap model of Zhang et al.\cite{zhang04} has incorporated the 
dependence of gap geometry on $\alpha$.  The extent of the
gap along the last open field line depends on $\alpha$, because its
inner radius, $r_{\rm in}$, which is assumed to be the null surface, 
is closer to the neutron star surface for oblique rotators.  The growth of the
gap across the field lines is limited by pair production on soft X-ray photons
that originate from the neutron star due to surface cooling, polar cap heating or 
outer gap heating.  Thus, the outer gap luminosity is expressed as
\be  \label{eq:LOG}
L_{OG}  = f^3 (\left\langle r \right\rangle ,P,B)^{} L_{sd}, 
\ee
where $L_{sd}$ is the spindown luminosity, $f$ is the size of the gap across field 
lines as a fraction of the total open
field line volume, and is dependent on the average emission radius in the gap,
$\left\langle r \right\rangle$, period $P$ and surface magnetic field $B$.  $f$ is
determined by the location of the pair formation front with respect to the last open field 
line, which is evaluated from the pair production condition
\be  \label{eq:Epp}
E_X E_\gamma  (1 - \cos \theta _{X\gamma }) = 2(mc^2 )^2 
\ee
where $E_X$ and $E_\gamma$ are the energy of the soft X-ray and $\gamma$-ray photon,
and $\theta _{X\gamma }$ is the angle between their propagation directions along 
the field lines.  
We use this version of the outer gap model for the population synthesis presented
in this paper.

Newer developments of the outer gap model include that of Takata et al.\cite{takata06}, 
who solve Poisson's equation in two dimensions to determine the geometry of the gap
both along and across field lines.  They find that the gap may extend below the
null surface in the presence of external currents, as in the 
one-dimensional gap model of Hirotani et al.\cite{Hir03}, and that the gap width grows
across field lines as the pulsar ages.  This picture of the outer gap begins to resemble
the slot gap in geometry, although the electrodynamics remains fundamentally different.
Cheng\cite{cheng06} has shown that this outer gap model may better reproduce light curves of 
known $\gamma$-ray pulsars than the original model.

\section{Radio Emission Model}
\label{sec:radmod}

Because the mechanism responsible for the radio beams is not understood, and more importantly 
because the radiation is coherent, it has not been possible to describe this emission using a physical 
model.  The emission has therefore been described using empirical models, developed over the years 
through detailed study of pulse morphology and polarization characteristics.  The average-pulse
profiles are quite stable and typically show a variety of shapes, ranging from a single 
peak to as many as five separate peaks.  The emission is also highly polarized, and displays changes in polarization position angle across the profile that often matches the swing expected for a sweep across 
the open field lines near the magnetic poles in the Rotating Vector Model\cite{RC69}.  
Rankin's\cite{Rankin93} study of pulse morphology concluded that pulsar radio emission can be 
characterized as having a core beam centered on the magnetic axis and one or more hollow cone beams outside of the core.  Although Rankin's model assumes that emission fills the core and cone beams, other studies\cite{LM88}
conclude that emission is patchy and only partially fills the core
and cone beam patterns.  

Arzoumanian et al. (ACC)\cite{ACC02} fit average-pulse profiles of a small collection of pulsars at 400 MHz to a core
and single cone beam model based on the work of Rankin.  The summed flux from the two components seen at angle $\theta$ to the magnetic field axis (modified by \cite{Gon04} to include frequency dependence $\nu$) is
\be  \label{eq:Stheta}
S(\theta, \nu ) = F_{\rm core} e^{ - \theta ^2 /\rho _{\rm core}^2 }  + 
F_{\rm cone} e^{ - (\theta  - \bar \theta )^2 /\omega _e^2 } 
\ee
where 
\be \label{eq:Fi}
F_i(\nu) = {-(1+\alpha_i) \over \nu}\left({\nu\over 50 {\rm MHz}}\right)^{\alpha_i+1}{L_i\over \Omega_i D^2}
\ee
and the index $i$ refers to the core or cone, $\alpha_i$ is the spectral index of the total angle-integrated flux, $L_i$ is the luminosity of
component and $D$ is the distance to the pulsar.  The width of the Gaussian describing the core beam is
\be  \label{eq:rhocone}
\rho _{\rm core}  = 1.5^{\circ} P^{ - 0.5} 
\ee
where $P$ is the pulsar period in seconds.
The annulus and width of the cone beam of Arzoumanian et al.\cite{ACC02}, with frequency dependence of
Mitra \& Deshpande\cite{MD99} are
\be  \label{eq:thetabarACC}
\bar \theta  = 1.4^{\circ} \left( {1 + \frac{{66}{\rm MHz}}{{\nu _{obs} }}} \right)P^{ - 0.5} 
\ee
\be  \label{eq:widannACC}
w_e = {\bar\theta \sqrt{\ln 2}\over 3}
\ee
The solid angles for the core and cone beams are
\be
\Omega_{\rm core} = \pi\rho _{\rm core}^2
\ee
\be
\Omega_{\rm cone} = 2\pi^{3/2} w_e\bar \theta
\ee
ACC found the ratio of core to cone peak flux to be 
\be  \label{eq:ACCratio}
r = \frac{{F_{\rm core} }}{{F_{\rm cone} }} = \frac{{20}}{{3P}}\left( {\frac{{\nu _{obs} }}{{400{\rm{MHz}}}}} \right)^{ - 0.5} 
\ee
where the frequency dependence has been added by \cite{Gon04}.  Thus in this model, $F_{\rm core} > F_{\rm cone}$ for
nearly all pulsars and the core beam is completely dominant for short-period pulsars.   

We have incorporated several modifications to the ACC model for use in our population synthesis.  The first change is to the ACC core-to-cone peak flux ratio resulting from the recent work of Gonthier et al.\cite{Gon06a},
who have carried out a study of 20 pulsars having three peaks in their average-pulse profiles, at three frequencies, 400, 600 and 1400 MHz.  They find a core-to-cone peak flux ratio 
\be  \label{eq:Gratio}
r = \frac{{F_{\rm core} }}{{F_{\rm cone} }} = \left\{ {\begin{array}{*{20}c}
   {25\,P^{1.3} \nu _{\rm GHz}^{ -0.9}, ~~~~~~~~~~~~ P < 0.7s}  \\
   {4\,P^{ - 1.8} \nu _{\rm GHz}^{ -0.9}, ~~~~~~~~~~~~ P > 0.7s}  \\
\end{array}} \right.
\ee
that is consistent with the core-to-cone peak flux ratio of ACC at periods above about 1 s, but predicts 
that pulsars with $P \lsim 0.05$ s are cone dominated.  Such a picture is supported by polarization 
observations of young pulsars.  Crawford et al.\cite{CMK01}\cite{CK03} 
measured polarization of a number of pulsars
younger than 100 kyr, finding that they possess a high degree of linear polarization and very little circular
polarization.  Since conal emission typically shows high degrees of linear polarization, this strongly indicates that the emission from these young pulsars comes from part of a cone beam.  
Johnston \& Weisberg\cite{JW06}, 
studying polarization of 14 pulsars younger than 75 kyr, also find high degrees of linear
polarization and flat position angle swings.  They concluded that the emission was
from a single wide cone beam, that core emission was weak or absent, and that the height of the 
cone emission is between 1\% and 10\% of the light cylinder radius.
A high emission altitude for young (fast) pulsars was also found by Kijak \& Gil\cite{KG03}, in their study
of average-pulse profile widths and their dependence on frequency.  Assuming that the edges of the pulse
are near the last open field line, they find an emission radius of
\be  \label{eq:rKG}
r_{\rm KG}  \approx 40\, \left({\dot P\over 10^{ - 15}{\rm s\,s^{-1}}}\right)^{0.07} P^{0.3} \nu _{GHz}^{ - 0.26} 
\ee
where $r_{\rm KG}$ is in units of stellar radius.
Our second modification to the ACC model is therefore to add the radius dependence of cone beam emission from
Eq. (\ref{eq:rKG}) above, so that Eq. (\ref{eq:thetabarACC}) and (\ref{eq:widannACC}) for the cone annulus and width
are changed to
\be \label{eq:thetabar}
\bar \theta  =  (1.-2.63\,\delta_w) \rho_{\rm cone}
\ee
\be  \label{eq:widann}
w_e = \delta_w \rho_{\rm cone}
\ee
where $\delta_w = 0.18$ \cite{Gon06a}, and
\be  \label{rhocone}
\rho_{\rm cone} = 1.24^{\circ}\,  r_{\rm KG}^{0.5}\, P^{- 0.5}
\ee

The luminosities of the core and cone components are then
\be  \label{eq:Lcc}
L_{\rm cone} = {L_{\rm radio}\over 1+(r/r_0)},   L_{\rm core} = {L_{\rm radio}\over 1+(r_0/r)},
\ee
where
\be   \label{eq:r0}
r_0 = {\Omega_{\rm cone}\over \Omega_{\rm core}}{(\alpha_{\rm core}+1)\over (\alpha_{\rm cone}+1)}
{1\over r}\left({\nu\over 50\,\rm MHz}\right)^{\alpha_{\rm core}-\alpha_{\rm cone}}
\ee
where $\alpha_{\rm core} = -1.96$ and $\alpha_{\rm cone} = -1.32$, and
\be   \label{eq:Lradio}
L_{\rm radio}  = 2.87 \times 10^{10}\,P^{ - 1.3} \dot P^{0.4}\, {\rm mJy\cdot kpc^2\cdot MHz} 
\ee
as modified from ACC, where $\dot P$ is in units of $1\,\rm s\,s^{-1}$.

These changes to the radio emission model result in a significant revision of the geometry of radio
beams in young and fast pulsars, from a narrow core beam emitted near the neutron star surface to a
wide cone beam emitted at relatively high altitude.  Such a revision produces greater visibility 
of the radio emission for fast pulsars and will have an influence on the predicted Geminga fraction.

\section{Pulsar Population Synthesis}
\label{sec:psrpop}

\subsection{Galactic neutron star evolution}
\label{sec:GalNS}

The pulsar population synthesis code consists of two main parts.  In the first part, neutron stars
start from birth locations in the Galaxy, with given initial $P$, $B$, $\alpha$, and space
velocity, at constant birth rate.  They are evolved through the Galactic potential and through their spin 
evolution to the present time.  In the second part, each pulsar is assigned a radio and $\gamma$-ray
flux and ``detected" by different radio and $\gamma$-ray surveys.  The first part of the
code is described in some detail by Gonthier et al.\cite{Gon06b}.  In this paper, we have used the same
set of evolved neutron stars to compute the numbers of radio-loud and radio-quiet pulsars expected
assuming either high-altitude slot gap or outer gap $\gamma$-ray flux and radio flux predicted by the
revised radio emission model described in Section (\ref{sec:radmod}).

\subsection{Emission geometry and luminosity}
\label{sec:geom}

\begin{figure*}
%\vskip 9cm
%\hskip -5.0cm
\includegraphics[width=17.5cm]{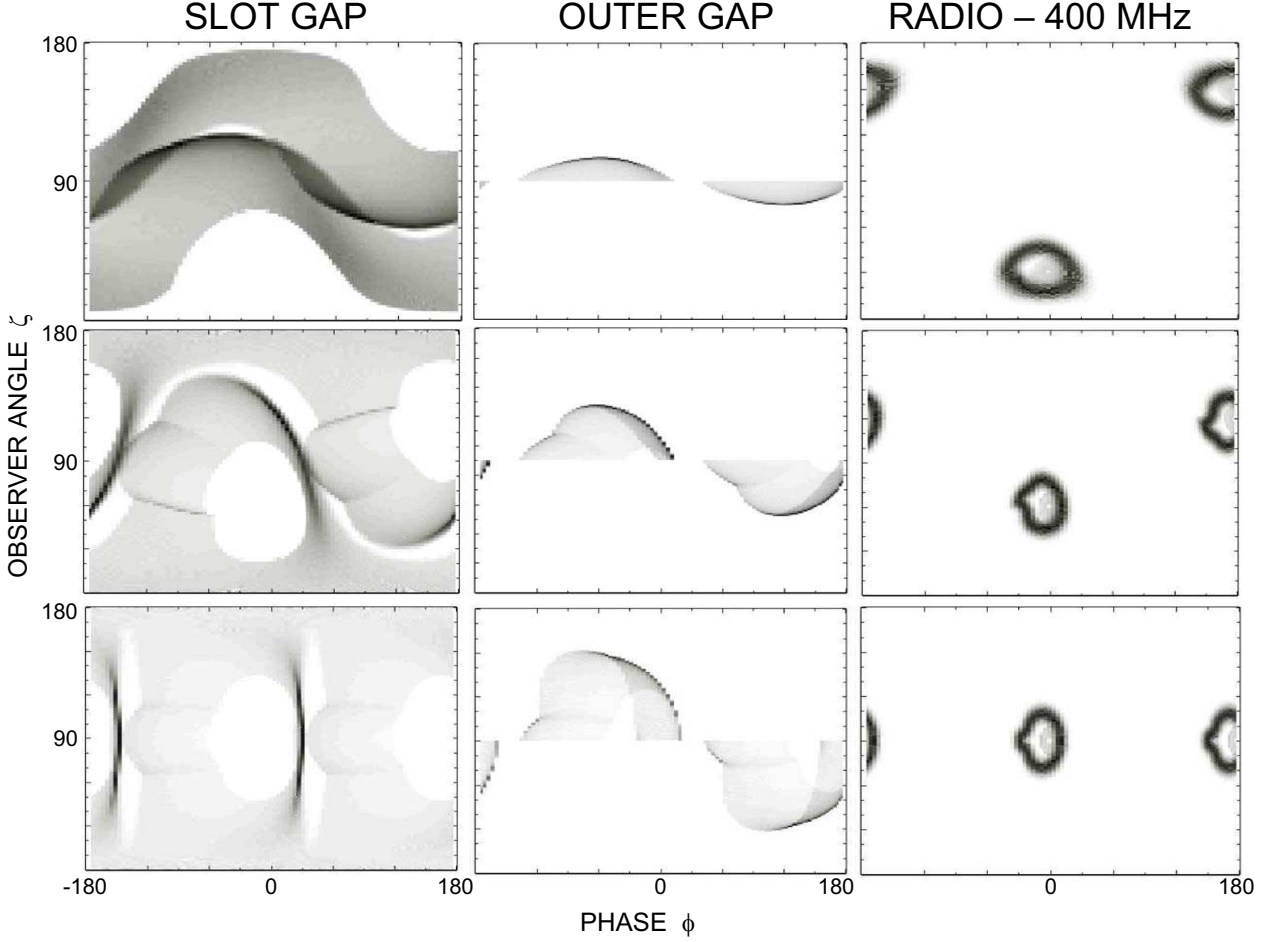}
\caption{Plots of emission in the ($\zeta$, $\phi$) plane for slot gap, outer gap and radio core and cone 
emission models, for $P = 50$ ms and $\alpha = 30^\circ$ (top), $60^\circ$ (middle) and $90^\circ$ (bottom).}
\label{fig:3}       % Give a unique label
\end{figure*}

To describe the emission geometry of each radiation model, we compute two-dimensional phase plots 
of emission in the ($\zeta$, $\phi$) plane, where $\zeta$ is the observer viewing 
angle measured from the rotation axis and $\phi$ is the rotational phase.  The method of computing the
phase plots is identical
to that of Dyks et al.\cite{DHR04}, who summed emission tangent to field lines of a retarded vacuum dipole,
taking into account relativistic effects of aberration and retardation.  The
emission was collected into bins on the sky (180 equal $\phi$ bins, 180 equal $\zeta$ bins), so that
each 2D element of the phase plot represents $dL/d\zeta d\phi$, where $dL = Id\Omega$ is differential 
luminosity and $I$ and $d\Omega$ are intensity and differential solid angle.
Sample phase plots for the slot gap, outer gap and radio cone and core beams are shown in Figure \ref{fig:3}.    
Caustics on the trailing field lines are evident in the slot-gap and outer-gap phase plots.  The shift of 
the cone beam relative to the core beam in radio phase plots is caused by aberration and retardation, as has 
been observed in radio profiles\cite{GG03,DRH04}.

Slot gap and outer gap phase 
plots are computed for nine values $\alpha$ equally spaced between $0^\circ$ and $90^\circ$.  In our
present calculation, the $\gamma$-ray phase plots depend only on $\alpha$ since we ignore any
dependence of the radiation pattern on pulsar age.  Radio phase plots are computed for the same values of 
$\alpha$, but also for core and cone components separately and for
five separate periods $P = 0.03, 0.05, 0.1, 0.5, 1.0$ s and two different frequencies, $\nu = 400$ MHz
and $\nu = 1400$ MHz.  The mean intensity $\bar I$ in each phase plot is then normalized to the total 
luminosity $2L$ from both poles divided by the total emission solid angle $2\Omega$ in each model, where
$\Omega$ is the solid angle of emission from each pole:
\be  \label{eq:Ibar}
\bar I = \left\langle {\frac{{dL}}{{d\Omega }}} \right\rangle  = \frac{{L }}{{\Omega }}
\ee
The phase-averaged flux,  
\be  \label{eq:nuFnu}
\left\langle {\nu F_\nu  } \right\rangle  = \frac{{\int {I(\alpha ,\varsigma ,\phi )d\phi } }}{{2\pi D^2 }}
\ee
is then determined by integrating the emission in the phase plot at a given $\zeta$, after 
interpolation in $\alpha$ (and in $P$ for the radio phase plots).  
This is equivalent to computing the phase-averaged flux in an observed pulse
profile.

For the slot gap model, we have assumed constant emission along the last open field line from the stellar 
surface to an altitude of $0.8\,R_{\rm L}$, where $R_{\rm L} = c/\Omega$ is the light cylinder radius.  
The total luminosity divided by solid angle from each pole is 
(from \cite{MH03})
\begin{eqnarray}  \label{eq:L_SG}
 \frac{L_{SG} }{\Omega _{SG}} = 
\varepsilon _\gamma \,\, [0.123\cos ^2 \alpha  + 0.82\,\theta _{PC}^2 \sin ^2 \alpha ]\,\, {\rm erg\,s^{-1}\,sr^{-1}}
\nonumber \\
\times \left\{
\begin{array}{ll}
9 \times 10^{34}\,  \left( {\frac{L_{sd}}{10^{35} {\rm erg/s}}} \right)^{3/7} P_{0.1}^{5/7}, & ~~~ B < 0.1\,B_{\rm cr} \\
2 \times 10^{34}\,  \left( {\frac{L_{sd}}{10^{35} {\rm erg/s}}} \right)^{4/7} P_{0.1}^{9/7}, & ~~~B > 0.1\,B_{\rm cr} 
\end{array}
\right.
\end{eqnarray}
where $P_{0.1} \equiv P/0.1$ s and we have assumed an efficiency of conversion of primary particle energy to $\gamma$-ray emission of $\varepsilon _\gamma   = 0.2$.  

For the outer gap model, we have assumed constant emission along the field line having magnetic colatitude
$\xi = \theta/\theta _{PC} = 0.85$, from the null surface to the light cylinder.  
The total luminosity of each pole is
$L_{OG}  = f^3 (\left\langle r \right\rangle ,P,B)^{} L_{sd}$, as given above by Eq. (\ref{eq:LOG}), and the solid angle of gap emission from each pole is (\cite{zhang04})
\be  \label{eq:Omog}
\Omega _{OG}  = 2\pi \left( {\frac{\alpha }{{90^\circ }}} \right)^2 \left( {\frac{{1 - bf}}{{1 + bf}}} \right)
\ee
where $b = <r>/R_L\sin\alpha$.
Using the above expressions we are able to reproduce the luminosity distribution in Fig. 2 of 
Zhang et al.\cite{zhang04}.

For the radio emission model, all the field lines contained in the open volume are used and the flux 
$S(\theta, \nu )$ from Eq (\ref{eq:Stheta}) gives the differential luminosity 
in each sky bin:
\be  \label{eq:L_r}
dL^i_\nu = D^2\, S_i(\theta, \nu )\sin\theta d\theta d\phi_{pc} d\nu 
\ee
emitted at altitude $1.8 R$ for the core component and at altitude given by Eq (\ref{eq:rKG}) for the conal
component, where $\phi_{pc}$ is the magnetic azimuth. The total
intensity of the phase plots are then normalized to $2\int\int S_i(\theta, \nu ) d\Omega D^2$.

\begin{figure*}
\centering
%\vskip 9cm
\includegraphics[height=4.5cm]{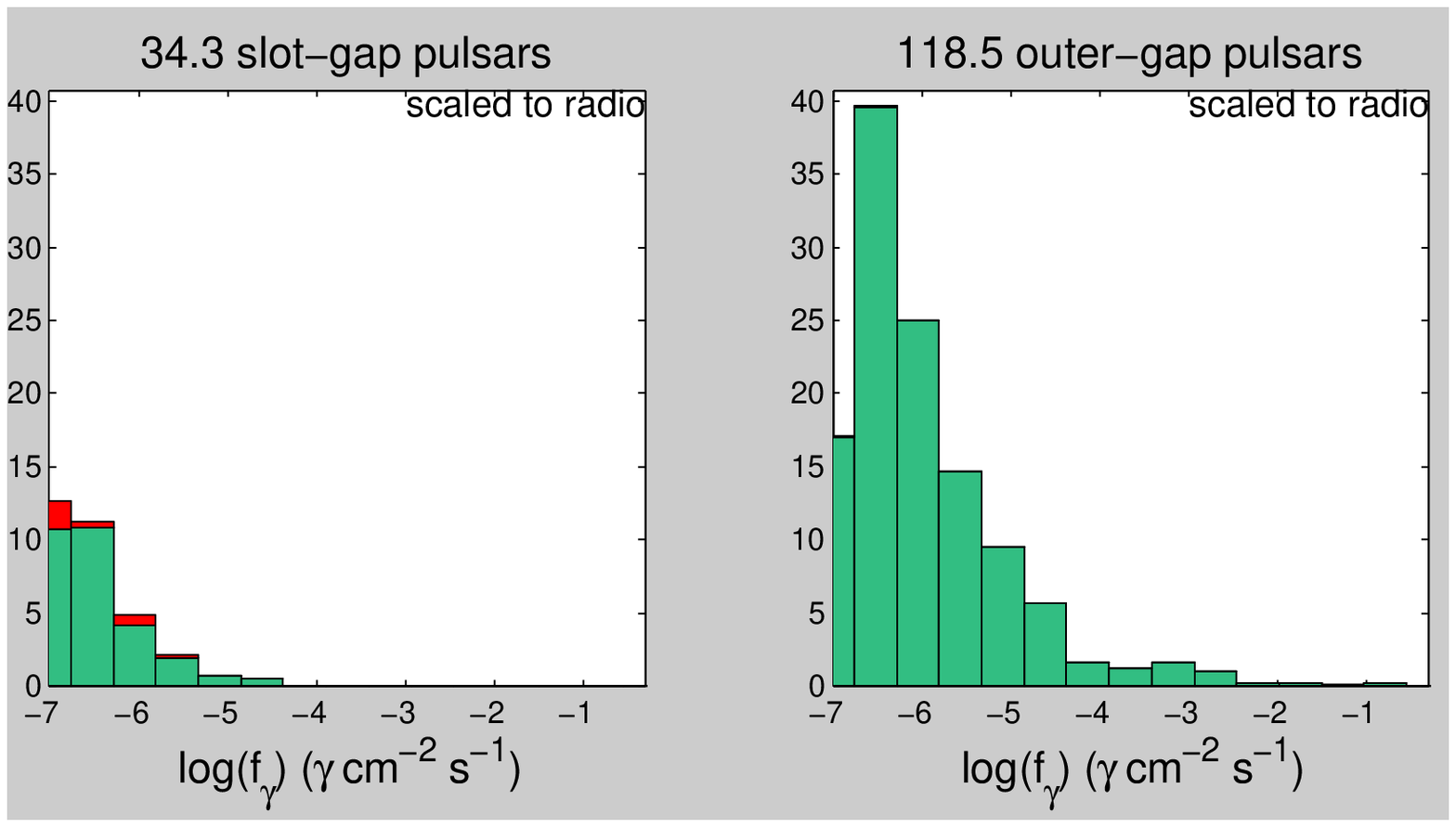}
\includegraphics[height=4.5cm]{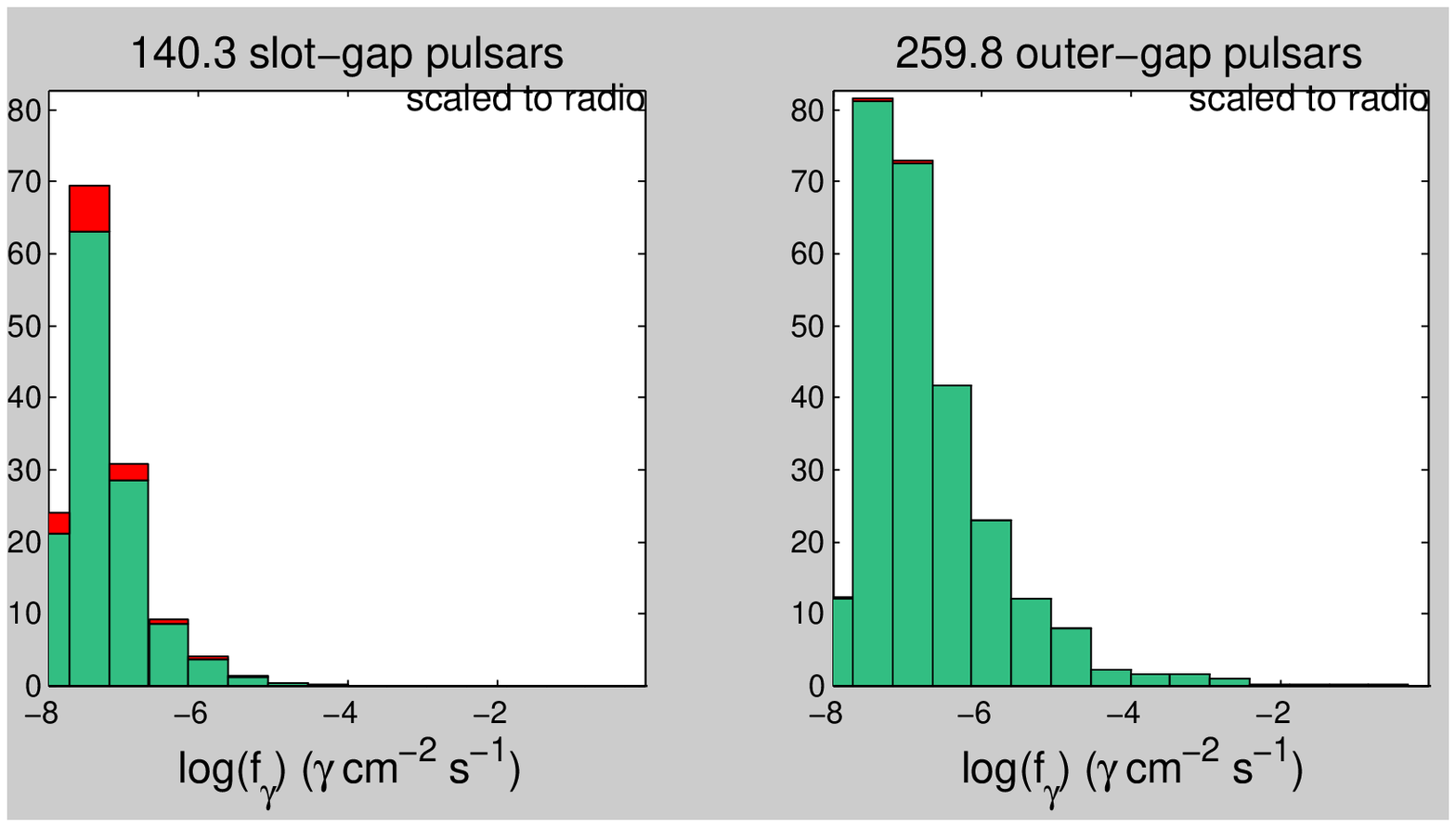}
\caption{Flux distribution of detected $\gamma$-ray pulsars for the slot gap and outer gap models for EGRET (left) and LAT (right). Radio-loud pulsars are in dark shading and radio-quiet pulsars are in light shading.}
\label{fig:4}       % Give a unique label
\end{figure*}

\section{Predicted Populations of Gamma-Ray Pulsars}
\label{sec:predpop}

\begin{figure*}
\centering
%\vskip 9cm
\includegraphics[height=4.5cm]{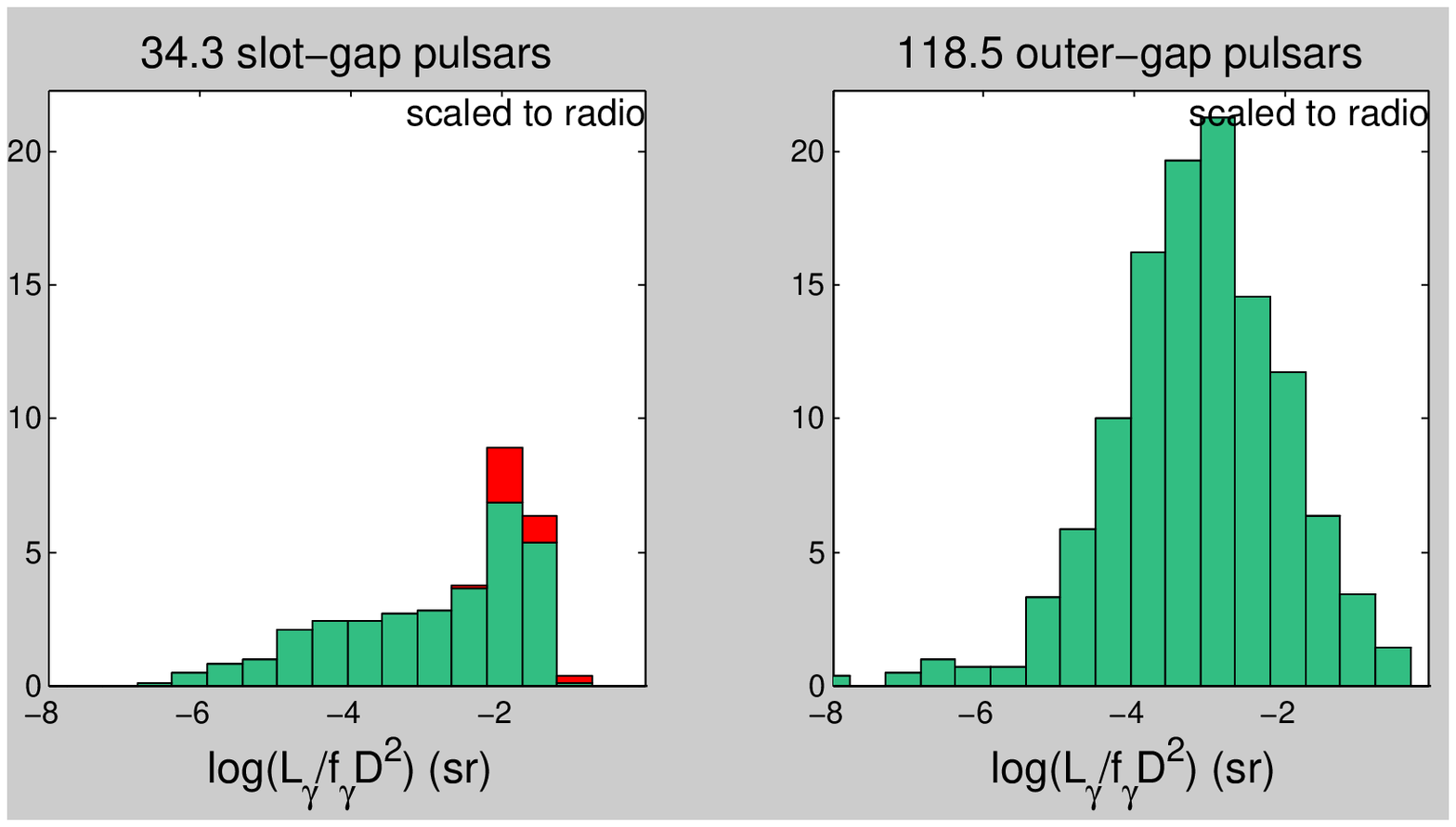}
\includegraphics[height=4.5cm]{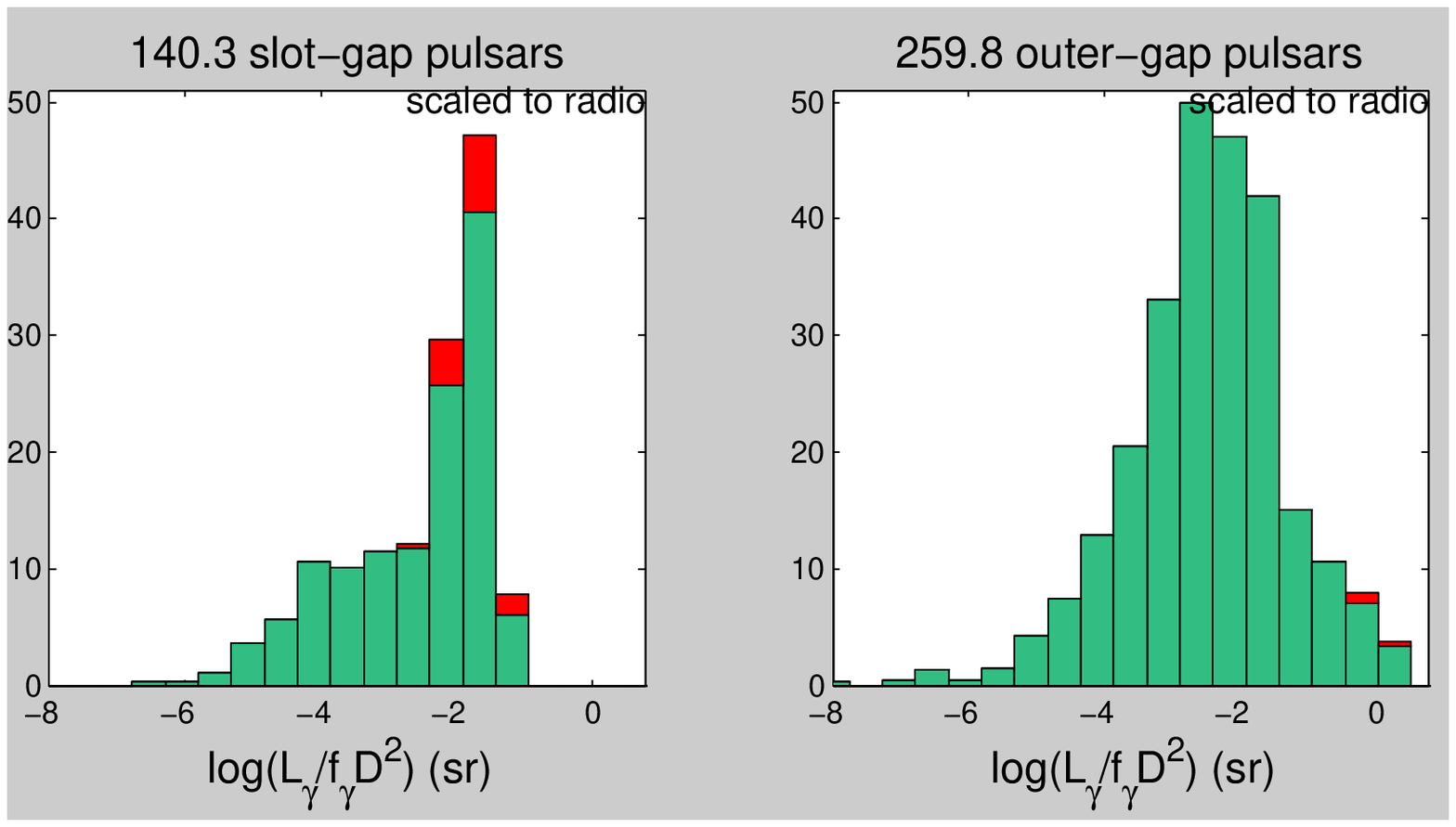}
\caption{log$(L_{\gamma}/F_\gamma D^2)$ distribution of detected $\gamma$-ray pulsars for the slot gap and outer gap 
models for EGRET (left) and LAT (right).}
\label{fig:5}       % Give a unique label
\end{figure*}

\begin{table*}[t]
\caption{Predicted numbers of $\gamma$-ray pulsars and Geminga fraction$^a$}
\centering
\label{tab:1}       % Give a unique label
\begin{tabular}{lcccccc}
\hline\noalign{\smallskip}
%\tableheadseprule\noalign{\smallskip} \hline
 & \multicolumn{4}{c}{Slot gap} & \multicolumn{2}{c}{Outer gap}  \\[3pt]
& \multicolumn{2}{c}{Low altitude$^b$} & \multicolumn{2}{c}{High altitude} && \\
\tableheadseprule\noalign{\smallskip} \hline
 Instrument & Radio-Loud & Radio-Quiet & Radio-Loud & Radio-Quiet & Radio-Loud & Radio-Quiet \\ \hline
EGRET & {\bf 25} & {\bf 10}~~ (0.28) & {\bf 3} & {\bf 31}~~ (0.91) & {\bf 0-1} & {\bf 118}~~ (0.99) \\
GLAST LAT & {\bf 94} & {\bf 51}~~ (0.35) & {\bf 13} & {\bf 128}~~ (0.91) & {\bf 1} & {\bf 258}~~ (0.99) \\ \hline
\noalign{\smallskip}\hline
$^a$in parentheses \\
$^b$From Gonthier et al.\cite{Gon06b}
\end{tabular}
\end{table*}

\begin{figure*}
\centering
%\vskip 9cm
\includegraphics[height=4.5cm]{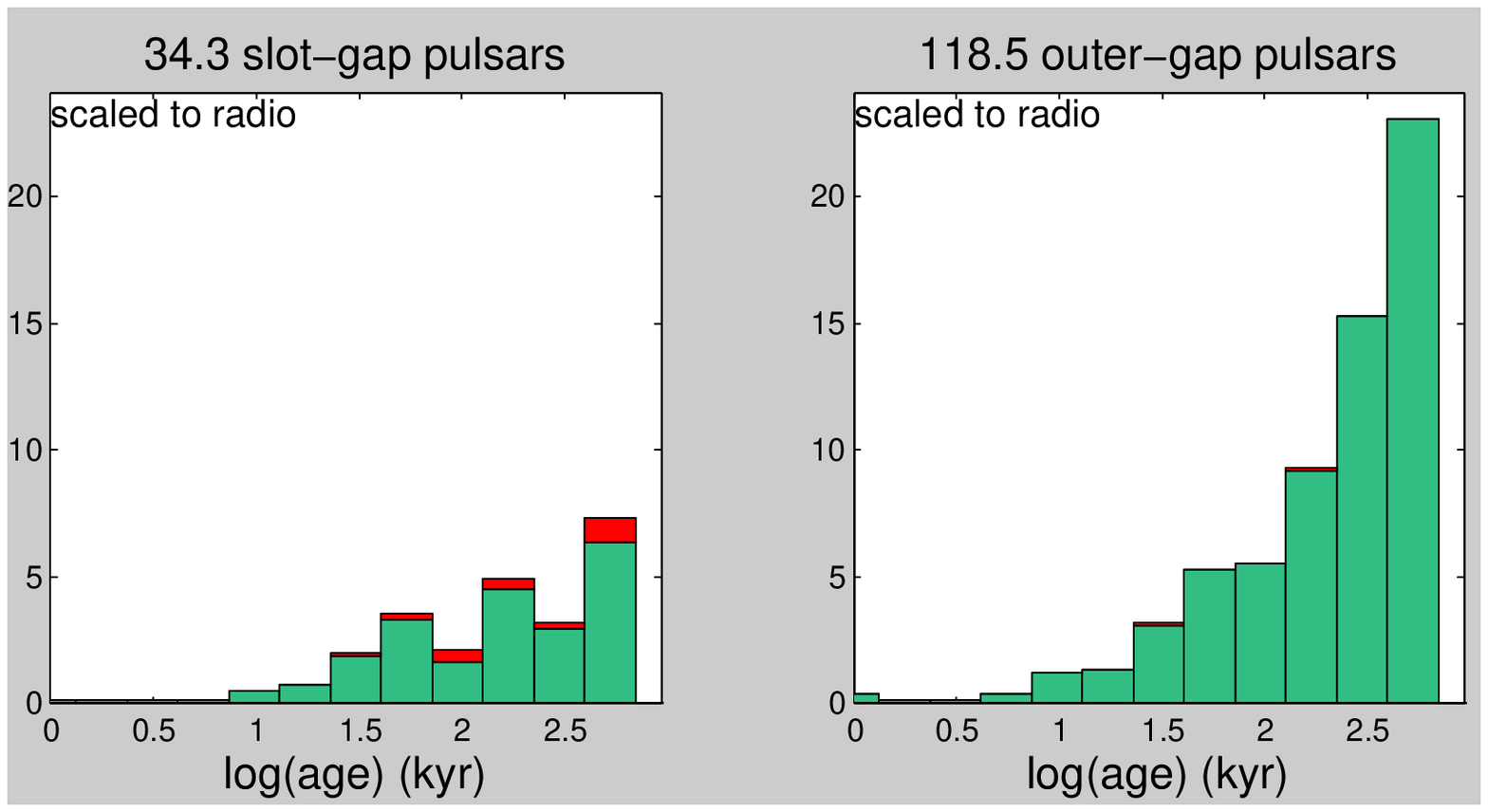}
\includegraphics[height=4.5cm]{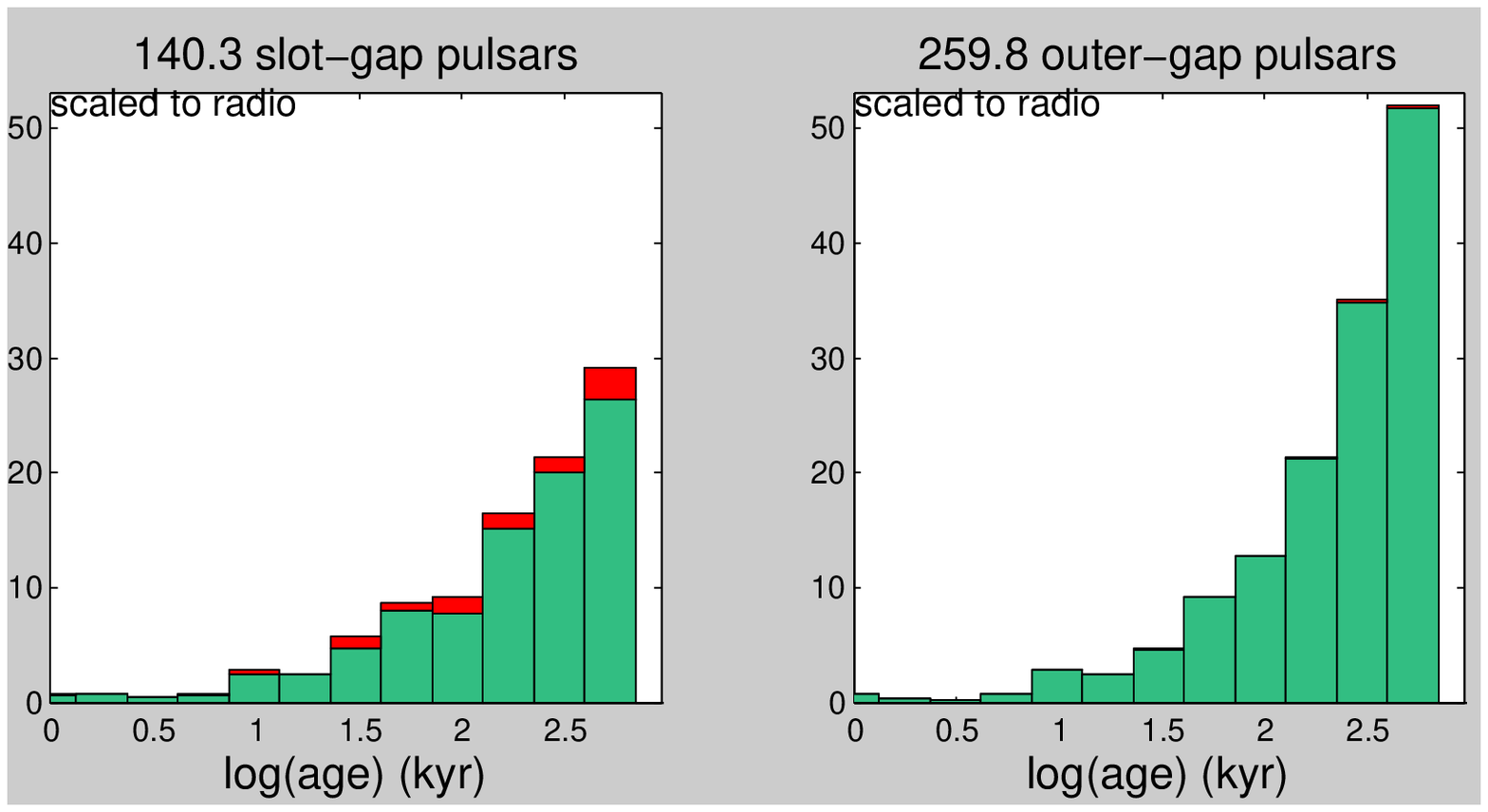}
\caption{log(age) distribution of detected $\gamma$-ray pulsars for the slot gap and outer gap models 
for EGRET (left) and LAT (right).}
\label{fig:6}       % Give a unique label
\end{figure*}

\begin{figure*}
\centering
%\vskip 9cm
\includegraphics[height=4.5cm]{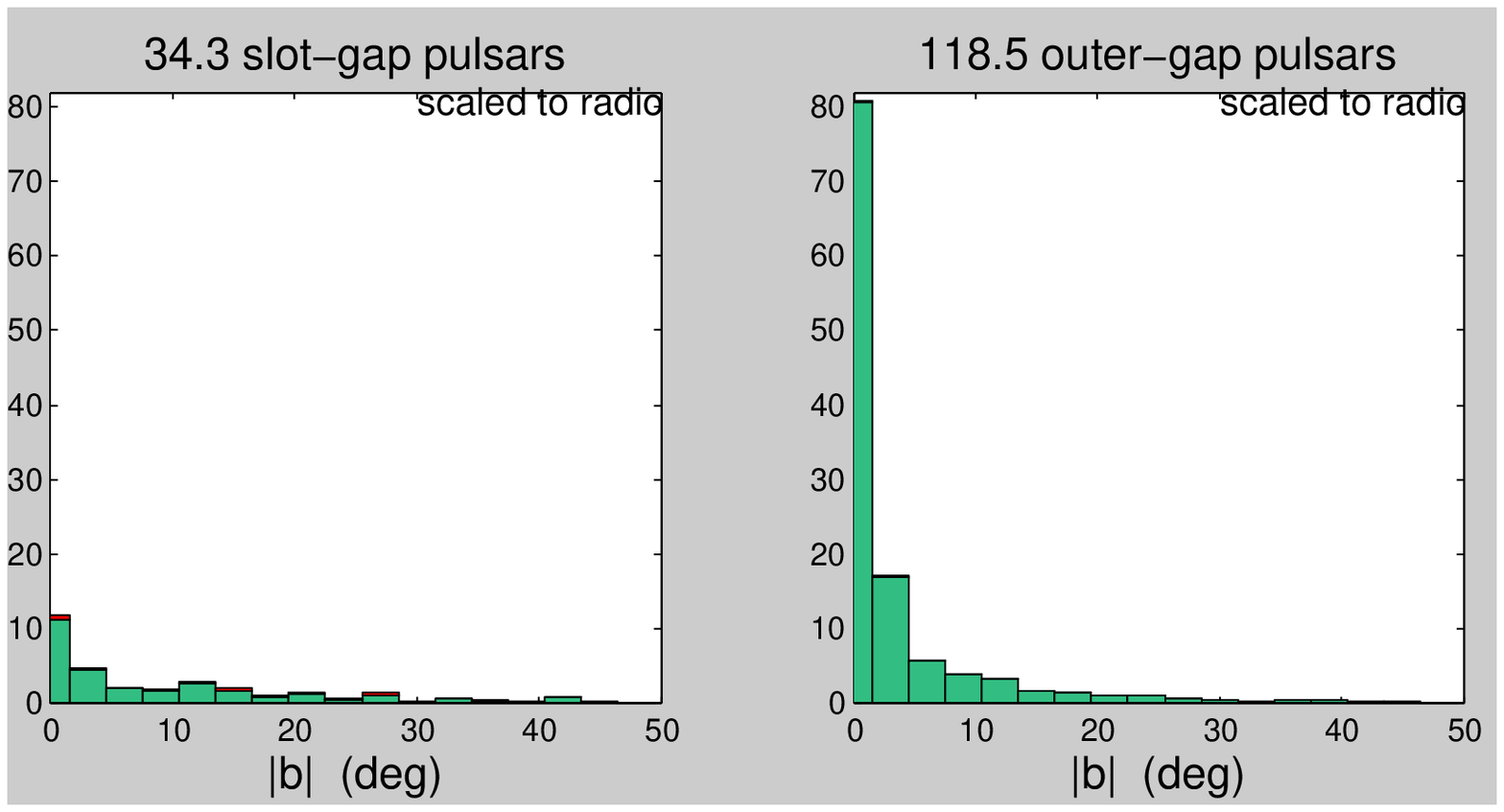}
\includegraphics[height=4.5cm]{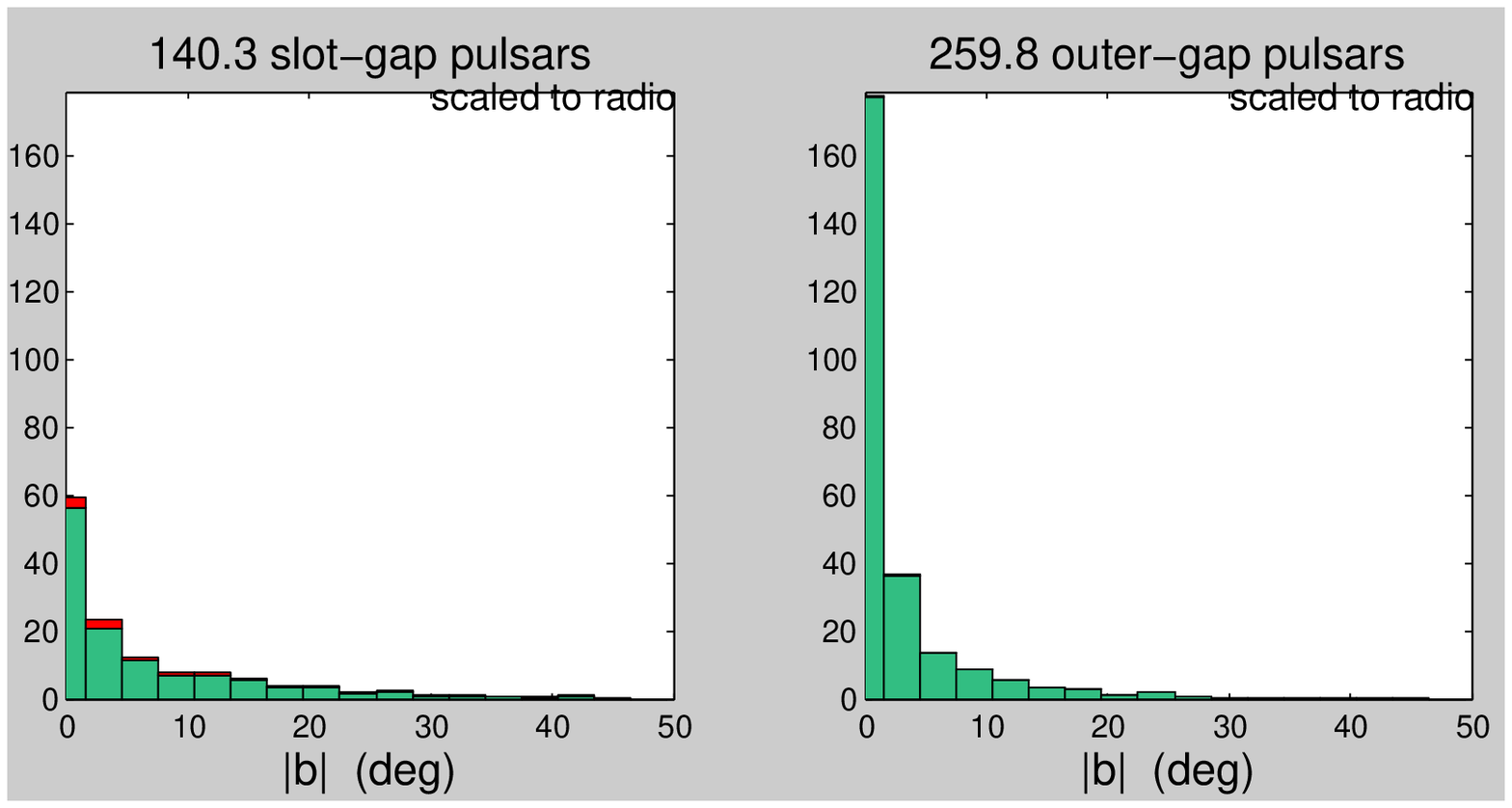}
\caption{Galactic latitude distribution of detected $\gamma$-ray pulsars for the slot gap and outer gap models 
for EGRET (left) and LAT (right).}
\label{fig:7}       % Give a unique label
\end{figure*}

Having assigned phase-averaged radio fluxes and $\gamma$-ray fluxes for either slot gap or outer gap models from 
the normalized phase plots to each evolved neutron star, we denote a source as radio loud if its radio flux
exceeds the thresholds of any of the ten surveys described in Gonthier et al.\cite{Gon06b}, and denote it as 
$\gamma$-ray loud if its $\gamma$-ray flux exceeds the flux thresholds of the 9 year EGRET or 
the GLAST 1 yr LAT (Large-Area Telescope) survey.
The number of simulated neutron stars is normalized by matching the number of detected radio pulsars to the 
total number detected by the ten surveys (which is equivalent to setting the neutron star birthrate). 
We use a revised EGRET threshold map, that takes into account the sky background model of Grenier et al.\cite{Grenier05},
and a GLAST 1 yr LAT threshold map.  In Table 1, we summarized the number of simulated radio-loud and 
radio-quiet $\gamma$-ray pulsars for slot gap and outer gap models predicted for EGRET and GLAST LAT telescopes,
as well as the resulting Geminga fraction in parentheses.
In both slot gap and outer gap models, the number of radio-quiet $\gamma$-ray pulsars far exceeds the number of
radio-loud $\gamma$-ray pulsars, thus predicting a large Geminga fraction.  Models where the high-energy emission
occurs mostly in the mid- or outer magnetosphere therefore produce many Geminga-like pulsars and very few radio-loud
pulsars.  The Geminga fraction in the outer gap model is significantly higher than in the slot gap model, with 
too few radio-loud pulsars to account for the number EGRET detected and only 1 predicted to be detectable in 1
year by the GLAST LAT, even though the total number of detectable $\gamma$-ray pulsars is a factor of 3 higher.  
The reason for the much higher number of $\gamma$-ray pulsars in the outer gap model is clear from Fig. \ref{fig:4}, which
shows the flux distribution of the pulsars detectable by both models.  It is evident that there are many more 
pulsars with high $\gamma$-ray fluxes in the outer gap model, although the distributions peak at about the same flux
in both models.  This is because although the luminosity distributions are similar, the outer gap pulsars have 
smaller effective solid angles, as shown in Figure \ref{fig:5}.  The age distributions of the detected EGRET
and LAT $\gamma$-ray pulsars, as shown in Figure \ref{fig:6}, are quite similar for the slot gap and outer gap
and the distribution strongly peaks at small ages.  However, the radio-loud pulsars appear to be somewhat more spread-out 
in age, although the statistics are limited.  The Galactic latitude distribution in Figure \ref{fig:7} 
shows that both the slot gap and outer gap $\gamma$-ray pulsars are concentrated near the Galactic plane, with 
the distribution of outer gap pulsars being much more strongly peaked at $|b| = 0$.  The $|b|$ distribution
of slot gap pulsars more closely resembles the latitude distribution of the EGRET sources.

Gonthier et al.\cite{Gon06b} have presented results for the numbers of radio-loud and radio-quiet pulsars with detectable 
emission from the low-altitude slot gap cascades, as well as from high-altitude slot gap emission.  
Those results are thus complimentary to these, and the results for the high-altitude slot gap can be compared.  
Both the high-altitude and low-altitude emission is expected to be present for pulsars having slot gaps, so the 
numbers in Table 1 should be combined to give the total number of $\gamma$-ray pulsars expected for the 
polar cap/slot gap model.  The Geminga
fraction for the low-altitude slot gap (0.28 for EGRET and 0.35 for GLAST) is much lower than for the high-altitude 
slot gap.  This is because the low-altitude slot gap cascade emission occurs at 3-4 stellar radii, much closer to 
the radio emission altitude for most of the $\gamma$-ray loud pulsars, especially the ones that are core dominated.
The low-altitude slot gap also produces enough EGRET radio-loud pulsars and can account for some fraction of the
unidentified $\gamma$-ray sources as radio-loud $\gamma$-ray pulsars, whereas we found that the 
outer magnetosphere models do not produce enough EGRET pulsars.

Jiang \& Zhang \cite{JZ06} have studied statistical properties of $\gamma$-ray pulsars in the outer gap model.
From their calculation, they predict that 8 radio-loud and 24 radio-quiet $\gamma$-ray pulsars are detectable 
by EGRET, and 78 radio-loud and 740 radio-quiet $\gamma$-ray pulsars are detectable by GLAST.  The 
predicted Geminga fractions are therefore 0.75 and 0.9 for EGRET and GLAST.  These fractions are significantly lower than our calculated Geminga fractions for the outer gap.  But there are at least two major differences 
in our calculations.  First, we have used more recent full-sky threshold maps for EGRET and the GLAST LAT,
taking into account the sky coverage of each telescope and detection above the intense interstellar background, 
which has the effect of reducing the number of
detected $\gamma$-ray pulsars.  Jiang \& Zhang \cite{JZ06} used only in-plane and 
out-of-plane sensitivities, which do not accurately account for the large variation in sensitivity with
Galactic coordinates.  Second, we have taken into account the relative spatial orientations of the radio and
$\gamma$-ray beams, as well as their solid angles, whereas Jiang \& Zhang have taken into account only
the different solid angles and beaming fractions of the radio and $\gamma$-ray emission.  This is 
important since the outer gap emission direction is generally at large angles to the radio emission 
direction, and there is a large region of phase space where radio and $\gamma$-ray beams are not visible to
the same observer, e.g. for $\alpha \lsim 40^\circ$.

\section{Conclusions}

We have determined the numbers of radio-loud and radio-quiet $\gamma$-ray pulsars using two different 
$\gamma$-ray models, from the same set of evolved Galactic neutron stars and using the same radio
emission model.  The radio emission model takes into account the most recent studies of radio pulse
emission morphology and polarization, which gives wide, relatively high altitude radio cone beams for 
the young $\gamma$-ray 
bright pulsars.  The full geometry of both radio and $\gamma$-ray emission is modeled, including 
relativistic effects of retardation and aberration, and distortion of the open field lines due to
a retarded dipole description.
The Geminga fraction is large for models such as the extended slot gap and outer gap where $\gamma$-ray emission 
occurs at high altitude in the pulsar magnetosphere, for a radio beam model that describes the bulk of
the radio pulsar population.  
Even the larger radio beams of young pulsars ($P \lsim 50$ ms)  
emitted at high altitude, produce few radio-loud pulsars and a low Geminga fraction,
since the number of fast pulsars ($P \lsim 50$ ms) with very large radio beams is not large enough.
For the bulk of the pulsars with $P \gsim 50$ ms, the radio beam size decreases rapidly while the $\gamma$-ray
beams remain large.  Both slot gap and outer gap models yield 
a large spread of the $f_\gamma D^2$ vs. $L_{SD}$ distribution.  In general, the radio-loud pulsars are closer
and have larger $L_{\gamma}/F_\gamma D^2$ than the radio-quiet pulsars.  
From the results of this paper for high-altitude
$\gamma$-ray emission models, combined with the results of Gonthier et al.\cite{Gon06b} in this proceedings for 
the polar cap/low-altitude slot gap model, we can come to the important conclusion that if many of the EGRET
unidentified sources are radio-loud $\gamma$-ray pulsars (and not pulsar wind nebulae), 
the $\gamma$-ray emission must come from relatively low
altitudes.  Of course, it is quite possible that different emission models may apply to pulsars of different
ages.  Future observations with GLAST of the number of radio-loud $\gamma$-ray pulsars, and a 
related limit on the Geminga fraction, will be able to distinguish between low-altitude and high-altitude
emission models.

%\begin{acknowledgements}
%\end{acknowledgements}

% BibTeX users please use
%\bibliographystyle{spmpsci}


\end{document}